\documentclass[10pt]{article}

\newcommand{\be}{\begin{equation}}
\newcommand{\ee}{\end{equation}}
\newcommand{\bea}{\begin{eqnarray}}
\newcommand{\eea}{\end{eqnarray}}

\newcommand{\R}{{\sf R\hspace*{-0.9ex}\rule{0.15ex}%
{1.5ex}\hspace*{0.9ex}}}
\def\bkR{{\rm I\kern-.17em R}}
\def\bkC{{\rm \kern.24em \vrule width.05em height1.4ex depth-.05ex \kern-.26em C}}
\def\bkN{{\rm \kern.50em \vrule width.05em height1.4ex depth-.05ex \kern-.26em N}}

\begin{document}

\markboth{Authors' Names}
{Instructions for Typing Manuscripts (Paper's Title)}

\author{Nuno Costa Dias\footnote{{\it ncdias@mail.telepac.pt}} \\ Jo\~{a}o Nuno Prata\footnote{{\it joao.prata@ulusofona.pt}} \\ {\it Departamento de Matem\'atica} \\
{\it Universidade Lus\'ofona de Humanidades e Tecnologias} \\ {\it Av. Campo Grande, 376, 1749-024 Lisboa, Portugal}}

\title{Deformation Quantization and Wigner Functions\footnote{{\it Talk presented by the second author at the Workshop on Quantum Gravity and Noncommutative Geometry, 20-23 July 2004, Universidade Lus\'ofona, Lisbon, Portugal.}}
}

\maketitle
\begin{abstract}
We review the Weyl-Wigner formulation of quantum mechanics in phase space. We discuss the concept of Narcowich-Wigner spectrum and use it to state necessary and sufficient conditions for a phase space function to be a Wigner distribution. Based on this formalism we analize the modifications introduced by the presence of boundaries. Finally, we discuss the concept of environment-induced decoherence in the context of the Weyl-Wigner approach.
   
\end{abstract}

\section{The Weyl-Wigner formulation of quantum mechanics}

The Wigner function was introduced in the 1930's \cite{Wigner}. Wigner was trying to derive quantum corrections to Boltzmann's kinetic equation. With this aim in mind he had to define some quantum counterpart to the Liouville density $\rho_{cl}$.

Let us recall some basic facts about this distribution. $\rho_{cl}$ is assumed to be a real, normalized, non-negative, $C^{\infty}$ function in phase-space. Here, we assume a flat $2n$-dimensional phase-space $T^* M \simeq \R^{2n}$. The expectation value of some observable $A \in C^{\infty} \left( T^* M; \R \right)$ is then given by:
\begin{equation}
<A> \bkN = \int dx \int dp \hspace{0.2 cm} A(x,p) \rho_{cl} (x,p).
\end{equation}
Moreover, the probability distribution that a measurement of $A$ yield the value $a \in \R$ is evaluated according to:
\begin{equation}
{\cal P} \left(A=a \right) = \int dx \int dp \hspace{0.2 cm} \delta \left( A (x,p) -a \right)\rho_{cl} (x,p).
\end{equation} 
Finally, the time evolution of both (1) and (2) can be computed in the Heisenberg or Schr\"odinger pictures:
\begin{equation}
\begin{array}{c}
<A (t)> = \int dx \int dp \hspace{0.2 cm} A(x,p,t) \rho_{cl} (x,p) = \int dx \int dp \hspace{0.2 cm} A(x,p) \rho_{cl} (x,p,t)\\
\\
{\cal P} \left(A(t) =a \right) = \int dx \int dp \hspace{0.2 cm} \delta \left( A (x,p,t) -a \right)\rho_{cl} (x,p) = \\
\\
= \int dx \int dp \hspace{0.2 cm} \delta \left( A (x,p) -a \right)\rho_{cl} (x,p,t)
\end{array}
\end{equation}
where $A(x,p,t)$ and $\rho_{cl} (x,p,t)$ are the solutions of Newton's and Liouville's equations, respectively: $\dot A = \left\{A,H \right\}_P$, $\dot{\rho}_{cl} = \left\{H , \rho_{cl} \right\}_P$.

Wigner's purpose was to establish some correspondence rule such that the expectation value of some hermitian operator $\hat A$ be given by:
\begin{equation}
<\hat A> = \int dx \int dp \hspace{0.2 cm} A(x,p) F (x,p),
\end{equation}
in analogy with classical statistical mechanics. The objects $A$ and $F$ would be the phase-space counterparts of $\hat A$ and of the density matrix $\hat{\rho}$, respectively. It is important to stress that the choice of $F$ and $A$ is not unique due to different possible choices of operator orderings. In this paper we shall adopt Wigner's choice, namely Weyl's rule \cite{Wigner}-\cite{Flato}. We define a $\bkC$-linear map - the Weyl map $W$ - which attributes to an element $\hat A$ of the quantum algebra of linear operators ${\cal A}_Q$ an element $A$ in the "classical" algebra ${\cal A}_C$ according to:
\begin{equation}
W \left( \hat A \right) = \left(\frac{\hbar}{2 \pi} \right)^n \int d \xi \int d \eta \hspace{0.2 cm} Tr \left\{\hat A \left( \hat q, \hat p \right) e^{- i \xi \hat q - i \eta \hat p} \right\} e^{i \xi x + i \eta p}.
\end{equation} 
This map appears naturally as a replacement $\left(\hat q, \hat p \right) \longleftrightarrow \left(x , p \right)$, once the operator has been written as a linear combination of completely symmetrized polynomials of $\hat q$ and $\hat p$. 

An important consequence of this map is that it introduces a noncommutative product (the starproduct \cite{Groenewold}) and a bracket (the Moyal bracket \cite{Moyal}) in phase space via the relations:
\begin{equation}
W \left( \hat A \cdot \hat B \right) \equiv A * B, \hspace{1 cm} W \left( \left[ \hat A , \hat B \right] \right) \equiv \frac{1}{i \hbar} \left[A, B \right]_M,
\end{equation}
where $A= W ( \hat A )$ and $B= W ( \hat B )$. From (5,6) one can obtain the explicit expressions:
\begin{equation}
\begin{array}{l}
 A * B  = A e^{\frac{i \hbar}{2} {\hat{\cal J}}} B = A \cdot B + {\cal O} \left( \hbar \right) \\
\\
\left[ A , B \right]_M  = \frac{2}{\hbar} A \sin \left(\frac{\hbar}{2} {\hat{\cal J}} \right) B = \left\{ A , B \right\}_P + {\cal O} \left( \hbar^2 \right) 
\end{array}
\end{equation}
where ${\hat{\cal J}}$ is the "{\it Poisson}" operator: $
{\hat{\cal J}} \equiv \left( \frac{ {\buildrel { \leftarrow}\over\partial}}{\partial q} \frac{ {\buildrel { 
\rightarrow}\over\partial}}{\partial p} -  \frac{{\buildrel { \leftarrow}\over\partial}}{\partial p}  \frac{{\buildrel { 
\rightarrow}\over\partial}}{\partial q} \right)  $, 
the derivatives ${\buildrel { \leftarrow}\over\partial}$ and ${\buildrel { \rightarrow}\over\partial}$ acting on $A$ and $B$, respectively. From (7) we realize that the starproduct and the Moyal bracket are $\hbar$-deformations of the usual commutative product of functions and of the Poisson bracket, respectively. The noncommutativity of the starproduct, on the other hand, is manifest, if we consider the relation $A *_{\hbar} B = B *_{- \hbar} A$. 

It is also interesting to remark the following property:
\begin{equation}
\int dx \int dp \hspace{0.2 cm} A(x,p) * B(x,p) = \int dx \int dp \hspace{0.2 cm} A(x,p) \cdot B(x,p).
\end{equation}
 
Thus far, we established the relation between the observable $\hat A$ and the phase space symbol $A = W( \hat A )$. Similarly, we have $F(x,p) = \left(2 \pi \hbar \right)^{-n} W \left( \hat{\rho} \right)$. This object is called the Wigner function and is the quantum mechanical counterpart of the classical Liouville measure $\rho_{cl}$. If the system is in a pure state with $\hat{\rho} = | \psi>< \psi |$, then the Wigner function reads:
\begin{equation}
F (x,p) = \frac{1}{\left(2 \pi \right)^n} \int d \eta \hspace{0.2 cm} e^{-i \eta p} \psi^* \left( x - \eta \hbar /2 \right) \psi \left( x + \eta \hbar /2 \right).
\end{equation}
It is worth summarizing some of the properties of Wigner functions. They are real, smooth and normalized as the classical Liouville measures $\rho_{cl}$. Likewise, expectation values of operators are computed according to the formula (4), reminescent of the classical expression (1). However, and contrary to $\rho_{cl}$, the Wigner function takes on negative values \cite{Bracken} as is illustrated by the following simple example. Let us consider, the simple harmonic oscillator, with Weyl symbol:
\begin{equation}
H(x,p) = \frac{p^2}{2m} + \frac{1}{2} m \omega^2 x^2.
\end{equation}
The Wigner function of the first excited state, then reads:
\begin{equation}
F_1 (x,p) = \frac{1}{\pi \hbar} \left( \frac{4 H (x,p)}{\omega \hbar} -1 \right) \exp \left( - \frac{2 H (x,p)}{\omega \hbar} \right).
\end{equation}
We conclude that $F_1 (x,p)$ is negative inside the ellipse: $H(x,p) < \frac{\omega \hbar}{4}$.

The fact that Wigner distributions are not, in general, positive defined is not surprising. One would otherwise be entitled to interpret them as true probability measures in phase-space, which in turn would violate Heisenberg's uncertainty principle. This is why quantum distributions in phase space are called "{\it quasi-distributions}". In any case, the marginal distributions are {\it bona fide} probability distributions. From (9) one gets:
\begin{equation}
\int dp \hspace{0.2 cm} F(x,p) = | \psi (x) |^2 \ge 0, \hspace{1 cm} \int dx \hspace{0.2 cm} F(x,p) = | \tilde{\psi} (p) |^2 \ge 0,
\end{equation}
where $\tilde{\psi} (p)$ is the Fourier transform of $\psi (x)$.
Moreover, if $F_1$, $F_2$ are the Wigner functions associated with the states $|\psi>$, $|\phi>$, respectively, then one gets:
\begin{equation}
\int dx \int dp \hspace{0.2 cm} F_1 (x,p) F_2 (x,p) = \frac{1}{\left(2 \pi \hbar \right)^n} |< \psi| \phi>|^2 \ge 0.
\end{equation}
The previous inequality remains valid for any pair of Wigner functions $F_1$, $F_2$ whether pure or mixed, and it is the basis of the probabilistic interpretation in quantum phase space. In particular, if $F_a (x,p)$ is the Wigner function associated with the eigenstate $|\psi_a>$ of some hermitian operator $\hat A$ with eigenvalue $a$, then the probability distribution that a measurement of $\hat A$ on a system in a state $F(x,p)$ yield the eigenvalue $a$ is given by\footnote{We assume for simplicity a nondegenerate, continuous spectrum. More general cases are addressed exhaustively in ref.\cite{Dias2}}:
\begin{equation}
{\cal P} \left( A=a \right) = \left(2 \pi \hbar \right)^n\int dx \int dp \hspace{0.2 cm} F(x,p) F_a (x,p).
\end{equation}
It is interesting to remark that, since the projector $|\psi_a><\psi_a|$ satisfies $\hat A |\psi_a><\psi_a| = |\psi_a><\psi_a| \hat A = a |\psi_a><\psi_a|$, then upon application of the Weyl map, one obtains \cite{Fairlie}:
\begin{equation}
A(x,p) * F_a (x,p) = F_a (x,p) * A(x,p) = a F_a (x,p).
\end{equation}
One calls this a stargenvalue equation and $F_a$ is said to be a stargenfunction of $A (x,p) = W ( \hat A )$ with eigenvalue $a$. This equation admits the formal solution \cite{Dias2}:
\begin{equation}
\left( 2 \pi \hbar \right)^n F_a (x,p) =  \delta_* \left( A (x,p) -a \right) \equiv  \frac{1}{2 \pi} \int dk \hspace{0.2 cm} e_*^{ik \left( A(x,p) -a \right)},
\end{equation}
where $e_*^B = 1 + B +  B*B/2 +  B*B*B/ \left(3! \right) + \cdots$ is the non-commutative exponential. The distribution $\delta_*$ is called the non-commutative delta function because of its similarities with the ordinary Dirac delta function. From (14) and (16), one gets:  
\begin{equation}
{\cal P} \left( A=a \right) = \int dx \int dp \hspace{0.2 cm} F(x,p) \delta_* \left( A (x,p) -a \right),
\end{equation}
which is the quantum mechanical analog of (2). 

We now turn to the dynamics of the system. The density matrix is a solution of the von Neumann equation: $i \hbar \partial \hat{\rho} / \partial t = \left[ \hat H, \hat{\rho} \right]$. If we multiply this equation by $\left( 2 \pi \hbar \right)^{-n}$ and apply the Weyl map, we obtain the Moyal equation:
\begin{equation}
\dot F (x,p,t) = \left[H (x,p) , F(x,p,t ) \right]_M = \left\{H (x,p) , F(x,p,t ) \right\}_P  
 + {\cal O} \left( \hbar^2 \right)
\end{equation}
which is a deformation of the classical Liouville equation.

\vspace{0.3 cm}
\noindent
In this work, we will apply the previous formalism to systems with boundaries (section 3) and to decohering systems interacting with a heat bath (section 4). In the latter situation we shall derive some known results using the concept of Narcowich-Wigner spectrum (which is discussed in section 2) and suggest some further developments.

\section{Narcowich-Wigner Spectrum}

So far, starting from ordinary quantum mechanics, we derived the necessary set of rules to make physical predictions in quantum phase space. But if Wigner quantum mechanics is to be a self-consistent autonomous formulation of quantum mechanics in its own right, we should be able to reverse the arguments, i.e. starting from Wigner quantum mechanics we should be able to re-derive Schr\"odinger quantum mechanics.

The obvious questions to pose are: (i) given some smooth, real and normalized phase space function $F(x,p)$ how do we know whether it is a {\it bona fide} phase space representative of a quantum state? (ii) if $F(x,p)$ represents a quantum state, how do we know whether it is a pure or a mixed state? 

To answer the first question, we start by stressing that not all functions $F(x,p)$ satisfying the aforementionned set of properties are Wigner functions. Take for instance \cite{Tatarskii},
\begin{equation}
F(x,p) = \frac{a \omega}{2 \pi} e^{-a H (x,p)},
\end{equation}
where $H(x,p)$ is given by (10) and $a$ is some constant with dimensions $\left( energy \right)^{-1}$. This function is $C^{\infty}$, normalized and real. However, the product of dispersions, $\Delta_2 x \cdot \Delta_2 p = \frac{1}{a \omega}$, violates Heisenberg's uncertainty principle whenever $a > \frac{2}{\hbar \omega}$, in which case it cannot represent a quantum state.

To state the set of necessary and sufficient conditions \cite{Dias1}, we first need to define the notion of pure state in quantum phase space. We say that a real, smooth, normalizable function $F(x,p)$ is a pure state whenever it satisfies the differential equation:
\begin{equation}
\frac{\partial^2}{\partial y^2} \ln {\cal Z} (x,y) = \left( \frac{\hbar}{2} \right)^2 \frac{\partial^2}{\partial x^2} \ln {\cal Z} (x,y),
\end{equation}
where ${\cal Z} (x,y) = \int dp \hspace{0.2 cm} e^{iyp} F (x,p)$. Alternatively, the previous equation can be replaced by the more algebraic condition:
\begin{equation}
F(x,p) * F (x, p) = \frac{1}{2 \pi \hbar} F(x,p).
\end{equation}
From the previous equation we obtain,
\begin{equation}
\int dx \int dp \hspace{0.2 cm} F^2 (x,p) =\frac{1}{ 2 \pi \hbar}.
\end{equation}
We thus construct the set ${\cal F}_{pure}$ of all pure states in quantum phase space. The space of all states ${\cal F}$ is the set of real, normalizable, smooth functions $F (x,p)$ such that:
\begin{equation}
\int dx \int dp \hspace{0.2 cm} F (x,p) F_{pure} (x,p) \ge 0, \hspace{1 cm} \forall F_{pure} \in {\cal F}_{pure}.
\end{equation}
States in phase space satisfy the inequality:
\begin{equation}
\int dx \int dp \hspace{0.2 cm} F^2 (x,p)  \le \frac{1}{ 2 \pi \hbar},
\end{equation}
the equality holding only for pure states (cf.(22)). Mixed states also admit an algebraic representation, of which (21) is the particular expression for pure states. They are positive elements of the algebra \cite{Dias1}, i.e. $F(x,p)$ belongs to ${\cal F}$ iff there exists a smooth, square integrable function $g(x,p)$ such that\footnote{The function $g(x,p)$ is defined up to $*$-multiplication by an "unitary" phase space function, i.e. $g(x,p)$ and $u (x,p) * g(x,p)$ yield the same state provided $\overline{u(x,p)} * u(x,p) =1$}:
\begin{equation}
F(x,p) = \overline{g (x,p)} * g (x,p).
\end{equation}
The set of necessary and sufficient conditions for $F( x,p)$ to be a Wigner function stated above can be rephrased in other terms more suited for certain applications. But first we introduce the notion of symplectic Fourier transform of some function $f(x,p)$:
\begin{equation}
\tilde f (a) = \int dz \hspace{0.2 cm} f (z) e ^{i \sigma (z,a)},
\end{equation}
where we used the compact notation $z^T= (x,p)$, $a^T= (u,v)$ and $\sigma (z,a)$ is the symplectic form: 
\begin{equation}
\sigma (z,a) = z^T J a = (x,p) \left( 
\begin{array}{c c}
0 & 1 \\
-1 & 0 
\end{array}
\right) \left(
\begin{array}{c}
u\\
v
\end{array}
\right) = xv -up.
\end{equation}
The inverse symplectic Fourier transform is:
\begin{equation}
f (z) = \frac{1}{(2 \pi)^{2n}} \int dz \hspace{0.2 cm} \tilde f (a) e ^{i \sigma (a ,z)}.
\end{equation}
A well known property of Fourier transforms is that they turn convolutions,
\begin{equation}
(f \star g ) (z) \equiv \int dz' \hspace{0.2 cm} f (z-z') g (z'),
\end{equation}
into products:
\begin{equation}
\widetilde{(f \star g)} (a) = \tilde f (a) \cdot \tilde g (a).
\end{equation}
We then state the following \cite{Narcowich1}

\vspace{0.5 cm}
\noindent
{\underline{\bf Definition 1:}} Let $\tilde f (a)$ be a continuous function on the dual of the phase space. Then $\tilde f$ will be termed of $\eta$-positive type if, for every $m \in \bkN$ and any set of $m$ points $\left\{a_1, \cdots, a_m \right\}$ in the dual of the phase space, the $m \times m$ matrix $M$ with entries
\begin{equation}
M_{jk} = \tilde f \left(a_j -a _k \right) e^{i \eta  \sigma \left( a_k , a _j \right) /2},
\end{equation}
is self-adjoint and non-negative.

\vspace{0.3 cm}
\noindent
We are now in a position to state the set of necessary and sufficient contions for $F$ to belong to ${\cal F}$ equivalent to the conditions (23) \cite{Werner}-\cite{Narcowich2}.

\vspace{0.5 cm}
\noindent
{\underline{\bf Definition 2 (KLM Conditions):}} A function $F(z)$ is a Wigner function iff its symplectic Fourier transform $\tilde F (a)$ satisfies the KLM (Kastler, Loupias, Miracle-Sole) conditions:

\vspace{0.3 cm}
\begin{tabular}{l l}
(i) & $\tilde F (0) =1$,\\
(ii) & $ \tilde F (a)$ is continuous and of $\hbar $-positive type.
\end{tabular}

\vspace{0.3 cm}
\noindent
Similar conditions also characterize classical states, i.e. the Liouville measures. We just need to replace $\hbar$-positive by $0$-positive type, which is equivalent to $F (z)$ being nonnegative everywhere in phase space (Bochner's theorem).

More generally, we shall denote by Narcowich-Wigner (NW) spectrum ${\cal W} \left(F \right)$ of the function $F(z)$ the set of real values $\eta$ such that $\tilde F$ is of $\eta$-positive type. In particular, if $F$ is a Wigner function, then $\hbar \in {\cal W} \left( F \right)$. Another important point is that if a function $F(z)$ is of $\eta ( \ne 0)$-positive type, $\tilde F (0) =1$ and $\tilde F (a)$ is continuous, then $F(z)$ is continuous, square integrable and vanishes at infinity. 

For $\eta > 0$ we can define an $\eta$-starproduct: $A*_{\eta}B \equiv A e^{i \eta  \hat{\cal J} /2} B$. It then follows that if $F$ is of $\eta$-positive type, then there exists some smooth square integrable function $g$ such that: $F(z) = \overline{g (z)} *_{\eta} g (z)$. Given the properties of the starproduct, we have: $F (z) = g (z) *_{- \eta} \overline{g (z) } = \overline{\left(\overline{ g (z) } \right)} *_{- \eta} \overline{g (z) }$. We conclude that if $\eta \in {\cal W} \left( F \right)$, then also $- \eta \in {\cal W} \left( F \right)$. Moreover, it can be shown that (for $\eta \ne 0$):
\begin{equation}
\int dz \hspace{0.2 cm} F^2 (z) \le \frac{1}{2 \pi | \eta|}.
\end{equation}
In summary, the best way to think about $\eta (>0)$-positive functions is to regard them as Wigner functions where $\hbar$ is replaced by $\eta$.

Now, suppose that $F(z)$ is a pure state Wigner function and that $\eta \in {\cal W} \left( F \right)$. From (22) and (32) we have: $\int dz \hspace{0.2 cm} F^2 (z) = \frac{1}{2 \pi \hbar} \le \frac{1}{2 \pi | \eta |}$. We have thus proved the following:

\vspace{0.5 cm}
\noindent
{\underline{\bf Lemma:}} If $F \in {\cal F}_{pure}$ then ${\cal W} \left( F \right) \subset \left[- \hbar, \hbar \right]$.

\vspace{0.5 cm}
\noindent
The previous lemma concerning the NW spectrum of a pure state can be further refined:

\vspace{0.5 cm}
\noindent
{\underline{\bf Theorem 1:}} Let $F \in {\cal F}_{pure}$. If $F$ is a Gaussian then ${\cal W} \left( F \right) = \left[- \hbar, \hbar \right]$, otherwise ${\cal W} \left( F \right) = \left\{- \hbar, \hbar \right\}$.
 
\vspace{0.5 cm}
\noindent
The proof of this theorem will be given elsewhere \cite{Dias4}. As an example, Narcowich \cite{Narcowich1} computed the NW spectrum of the Gaussian,
\begin{equation}
F(z) = \sqrt{\frac{\det A}{\pi^{2n}}} \exp \left[ - \left( z- z_0 \right)^T A \left( z- z_0 \right) \right],
\end{equation}
where $A$ is a real, symmetric, positive defined $2n \times 2n$ matrix. It was shown that $\tilde F$ is of $\eta$-positive type iff the matrix $B = A^{-1} + i \eta J$ is non-negative. In particular, if $A= c^{-1} Id$ (in suitable units), where $c$ is some real positive constant, then ${\cal W} \left( F \right) = \left[- c, c \right]$. 

Another interesting property of NW spectra stems from the convolution of distributions. In fact, let the Schur product of two matrices $A$ and $B$ be the matrix $M$ with entries $M_{jk} = A_{jk} B_{jk}$. Then it can be shown that if $A$ and $B$ are non-negative, then so is $M$. From the definition of NW spectrum, we can prove the following \cite{Narcowich1}: 

\vspace{0.5 cm}
\noindent
{\underline{\bf Theorem 2:}} Let ${\cal W} \left( F \right) $, ${\cal W} \left( G \right)$ and ${\cal W} \left( F \star G \right)$ be the NW spectra of $F$, $G$ and of the convolution $F \star G$. Then we have: ${\cal W} \left( F \right)  + {\cal W} \left( G \right) \subseteq {\cal W} \left( F \star G \right)$.

\vspace{0.5 cm}
\noindent
In the previous relation, the set on the left-hand side is ${\cal W} \left( F \right)  + {\cal W} \left( G \right) = \left\{\left. \eta + \chi \right| \eta \in {\cal W} \left( F \right), \chi \in {\cal W} \left( G \right) \right\}$.

We now consider a simple application of the concept of the NW spectrum. Let $F_0 (z)$ be some real, continuous function with non-empty NW spectrum ${\cal W} \left( F_0 \right)$. What are the set of conditions that $F_0$ should satisfy so that its convolution with any Wigner distribution $F \in {\cal F}$ yield another Wigner function? Using the properties of NW spectra it is easy to derive a set of sufficient conditions. Whether they are also necessary remains unproved. The conditions are as follows \cite{Narcowich1}:

\vspace{0.5 cm}
\noindent
(i) Let $F_0$ be of $0$-positive type. If $F \in {\cal F}$, then $\pm \hbar \in {\cal W} \left( F \right)$. From theorem 2, we conclude that $\left\{- \hbar, \hbar \right\} \subseteq {\cal W} \left( F_0 \star F \right)$. Consequently, $F_0 \star F$ is a Wigner function for any $F \in {\cal F}$.

\vspace{0.5 cm}
\noindent
(ii) Let $\pm 2 \hbar \in {\cal W} \left( F_0 \right)$ and let $F \in {\cal F}$. Similarly, from theorem 2, we obtain $\left\{\pm \hbar, \pm 3 \hbar \right\} \subseteq {\cal W} \left( F_0 %
\star F \right)$. Again, $F_0 \star F$ is a Wigner function for any $F \in {\cal F}$.

\vspace{0.3 cm}
\noindent
These results have two interesting applications. First, suppose that $F_0$ is also a Wigner function $\left( \pm \hbar \in {\cal W} \left( F_0 \right) \right)$. We conclude that $F_0 \star F$ is of $0$-positive type for any $F \in {\cal F}$. This means that under conditions (i) and (ii), $F_0 \star F$ is a non-negative Wigner function. This has an interesting physical interpretation. If $F_0$ is taken to be a state associated with an apparatus, then $F_0 \star F$ can be regarded as being the state that results when one uses that apparatus to make a measurement on a system in the state $F$ \cite{Narcowich1}. The second application will be postponned to section 4.

\section{Wigner Functions with Boundaries}

Consider a particle of mass $m$ in the infinite potential well \cite{Dias3}, confined to the interval $a \le x \le b$. For simplicity we shall assume Dirichlet boundary conditions. In ordinary quantum mechanics we solve the free Schr\"odinger equation and impose the boundary conditions thereafter. The normalized solutions are $\left(a= -b = - L/2 \right)$:
\begin{equation}
\psi_n (x) = \sqrt{2/ L} \sin \left[\left(n \pi /L \right) \left( x + L /2 \right) \right], 
\end{equation} 
inside the box and $n=1,2,3, \cdots$ The energy eigenvalues are $E_n = \hbar^2 \pi^2 n ^2 / \left( 2m L^2 \right)$. If we compute the associated Wigner function using (9), we obtain for $- L/2 < x < L/2$:
\begin{equation}
\begin{array}{c}
F_n (x,p) =  \frac{(-1)^{n+1}}{\pi L} \cos \left(\frac{2n \pi x}{L} \right) \cdot \sin \left[\frac{2p}{\hbar} \left(\frac{L}{2} - |x| \right) \right] + \\
\\
+ \frac{1}{2 \pi \left( Lp + \hbar n \pi \right)} \sin \left[\frac{2}{\hbar L} \left( pL + \hbar n \pi \right) \left( \frac{L}{2} - |x| \right) \right] + \\
\\
+ \frac{1}{2 \pi \left( Lp - \hbar n \pi \right)} \sin \left[\frac{2}{\hbar L} \left( pL - \hbar n \pi \right) \left( \frac{L}{2} - |x| \right) \right]
\end{array}
\end{equation}
It is easy to verify that this expression satisfies Dirichlet boundary conditions $F_n \left( - L/2 ,p \right)= F_n \left( L/2 ,p \right) =0, \hspace{0.3 cm} \forall p \in \R$. However, and contrary to the wavefunction, it also obeys Neumann boundary conditions $\partial F_n / \partial x \left( - L/2 ,p \right) = \partial F_n / \partial x \left( L/2 ,p \right)= 0$ and also $\partial^2 F_n / \partial x^2 \left( - L/2 ,p \right) = \partial^2 F_n / \partial x^2 \left( L/2 ,p \right)= 0, \hspace{0.3 cm} \forall p \in \R$. But the most important aspect of this problem is the following. We expect the free particle in phase space to obey the stargenvalue equations:
\begin{equation}
\frac{p^2}{2m} * F_n (x,p) = F_n (x,p) * \frac{p^2}{2m} = E_n F_n (x,p),
\end{equation}
with the same eigenvalue $E_n$. However, a straightforward calculation shows that $F_n$ (35) is not a solution of (36). This may be linked to the non-local nature of the $*$-product, which entails that the presence of the boundaries is felt well inside the bulk. Also notice that since $e^{i \hat p/ \hbar}$ implements spatial translations and translation invariance is broken when there is a boundary, we conclude that the Heisenberg algebra is no longer valid. Remember that the Heisenberg algebra is at the heart of the Weyl quantization.

So the problem we want to solve is the following\footnote{For an alternative recent approach we refer to \cite{Walton}.}:

\vspace{0.5 cm}
\noindent
(i) How do we modify eq.(36) to accommodate the boundary effects?

\noindent
(ii) How do we modify the Moyal equation for the dynamics, when a boundary is present?

\noindent
(iii) How do we choose the boundary conditions satisfied by the Wigner function? 

\vspace{0.3 cm}
\noindent
First of all let us go back to the definition of the Wigner function (9). In the confined case, $\psi^* (x-y) \psi (x+y)$ vanishes outside of the parallelogram $a<x-y<b$, $a<x+y<b$. If $x_0= (a+b)/2$, then $F(x,p)$ can be defined sectionwise, for $a<x \le x_0$:
\begin{equation}
F(x,p) = F_1 (x,p ) \equiv \frac{1}{\pi \hbar} \int_{a-x}^{x-a} dy \hspace{0.2 cm} e^{-2ipy/ \hbar}  \psi^* (x-y) \psi (x+y),
\end{equation}
and for $x_0 < x <b$:
\begin{equation}
F(x,p) = F_2 (x,p ) \equiv \frac{1}{\pi \hbar} \int_{x-b}^{b-x} dy \hspace{0.2 cm} e^{-2ipy/ \hbar}  \psi^* (x-y) \psi (x+y).
\end{equation}
Since the wavefunction or its derivatives are possibly discontinuous at the boundaries and to avoid possible misinterpretations, the previous integrals are defined as improper: $\int_{a-x}^{x-a}$ stands for $\lim_{c \to a^+} \int_{c-x}^{x-c}$ and, likewise, $\int_{x-b}^{b-x}$ stands for $\lim_{c \to b^-} \int_{x-c}^{c-x}$. From definitions (37,38) it is easy to check that $F$ is continuous at $x=x_0$ and satisfies Dirichlet boundary conditions irrespective of the boundary conditions satisfied by the wavefunction. The boundary conditions on $\psi$ do play a role in the derivatives of $F$, namely: $\partial F / \partial x (a^+,p) = 2 |\psi (a^+ )|^2 / (\pi \hbar)$ and $\partial F / \partial x (b^-,p) = - 2 |\psi (b^- )|^2 / (\pi \hbar)$. And so, if $\psi$ obeys Dirichlet boundary conditions, then:
\begin{equation}
\frac{\partial F}{\partial x} (a^+,p) = \frac{\partial F}{\partial x} (b^-,p) =0, \hspace{1 cm}
\frac{\partial F}{\partial x} (x_0^-,p) = \frac{\partial F}{\partial x} (x_0^+,p),
\end{equation}
but also:
\begin{equation}
\frac{\partial^2 F}{\partial x^2} (a^+,p) = \frac{\partial^2 F}{\partial x^2} (b^-,p) =0, \hspace{1 cm}
\frac{\partial^2 F}{\partial x^2} (x_0^-,p) = \frac{\partial^2 F}{\partial x^2} (x_0^+,p),
\end{equation}
However, $\frac{\partial^3 F}{\partial x^3} (a^+,p) = 8 |\psi' (a^+) |^2 / (\pi \hbar) \ne 0$.

We are now in a position to compute the boundary corrections to the stargenvalue equations (36)\cite{Dias3}:

\vspace{0.5 cm}
\noindent
{\underline{\bf Theorem 3:}} If the wavefunction $\psi$ is a solution of the Shr\"odinger equation
\begin{equation}
- \hbar^2 /(2m) \psi'' (x) + V(x) \psi (x) = E \psi (x),
\end{equation}
with Dirichlet boundary conditions, then the associated Wigner function $F(x,p)$ solves\footnote{The infinitesimal parameter $\epsilon$ is to be taken to $0^+$ once the starproducts and the integrations have been evaluated.}:
\begin{equation}
\begin{array}{c}
\left( \frac{p^2}{2m} + V(x) \right) * F(x,p)   - \frac{\hbar^2}{2m} \delta' (x-a) * F(x+ \epsilon ,p) + \\
\\
 +  \frac{\hbar^2}{2m} \delta' (x-b)* F(x- \epsilon ,p)= E F(x,p).
\end{array}
\end{equation}
Moreover $F(x,p)$ obeys the boundary conditions:
\begin{equation}
 \int dp \hspace{0.2 cm} F (a + \epsilon,p) =  \int dp \hspace{0.2 cm} F (b - \epsilon,p) = 0.
\end{equation}

\vspace{0.5 cm}
\noindent
{\underline{\bf Proof:}} A straightforward calculation shows that (for $a < x < x_0$):
\begin{equation}
\left( \frac{p^2}{2m} + V(x) \right) * F_1(x,p) = E F_1(x,p) + {\cal B} (x,p),
\end{equation}
where the boundary contribution is:
\begin{equation}
\begin{array}{c}
{\cal B} (x,p) = - \frac{\hbar}{2 \pi m} e^{- 2i p (a-x) / \hbar} \left\{ \frac{2ip}{\hbar} \psi^* (2x- a^- ) \psi (a^+) + \right. \\
\\
\left. + \psi'^* (2 x -a^-) \psi (a^+) + \psi^* (2x- a^-) \psi' (a^+) \right\}.
\end{array}
\end{equation}
For Dirichlet boundary conditions, the previous equation reads:
\begin{equation}
{\cal B}^D (x,p) = - \frac{\hbar}{2 \pi m} e^{- 2i p (a-x) / \hbar}  \psi^* (2x- a^-) \psi' (a^+).
\end{equation}
A simple calculation shows that this can be expressed in terms of the Wigner function as:
\begin{equation}
{\cal B}^D (x,p) = \lim_{\epsilon \to 0^+} \frac{i \hbar^2}{4 \pi m} \int dk \hspace{0.2 cm} e^{ik (x-a)} k F_1 \left( x + \epsilon , p - \hbar k /2 \right).
\end{equation}
The parameter $\epsilon$ appears in the previous expression because we want the derivative $\psi' (a^+)$ and not $\psi' (a^-)$ (remember that $\psi'$ is discontinuous at $a$). If we expand ${\cal B}^D$ in powers of $\hbar$, we get:
\begin{equation}
{\cal B}^D (x,p) =   \frac{\hbar^2}{2m} \lim_{\epsilon \to 0^+} \delta' (x-a) * F(x+ \epsilon ,p) .
\end{equation}
Performing a similar calculation for $F_2$ in $x_0 < x < b$ and assembling all the results we obtain (42). That $F$ obeys the set of boundary  conditions (43) is a trivial consequence of the relation $\lim_{\epsilon \to 0^+} \int dp \hspace{0.1 cm} F \left (x \pm \epsilon , p \right) = | \psi \left(x^{\pm} \right)|^2 $.

\vspace{0.5 cm}
\noindent
Some remarks are now in order. First of all, it can be checked that eq.(35) is a solution of the boundary stargenvalue equation (42). Secondly, we can perform Baker's converse construction. This means the following. Suppose that some real, normalized function $F$ is solution a of eq.(42) and that it satisfies the boundary conditions (43). Then, we can show that there exists a square integrable function $\psi$ (unique up to a phase) which (i) obeys Dirichlet boundary conditions at $a$ and $b$, (ii) is a solution of the Schr\"odinger equation (41) and (iii) is related to $F(x,p)$ via the Weyl transform (9). Thirdly, a similar calculation for the dynamics yields the following boundary Moyal equation:
\begin{equation}
\begin{array}{c}
\frac{\partial F}{\partial t} (x,p,t) = \left[ \frac{p^2}{2m} + V (x) , F(x,p,t) \right]_M  + \\
\\
+ \frac{\hbar^2}{2m} \left[ \delta' (x-b), F (x - \epsilon , p , t) \right]_M  - \frac{\hbar^2}{2m} \left[ \delta' (x-a), F (x + \epsilon , p , t) \right]_M,
\end{array}
\end{equation}
as expected.

\section{Decoherence}

The paradigm of environment induced decoherence consists of a Brownian particle interacting with a heat bath. The Brownian particle may be regarded as an open system, the behaviour of which will usually be described (under general conditions) by master equations with a local time. In a typical situation \cite{Diosi} of a dust particle interacting with air molecules or radiation, the dust particle can then be described by a Markowian master equation for the reduced density matrix\footnote{This is the operator which results when the environment's degrees of freedom have been traced out from the closed system's density matrix.}:
\begin{equation}
\frac{d \hat{\rho}}{dt} = - \frac{i}{2m \hbar} \left[\hat p^2, \hat{\rho} \right] - \frac{D}{2} \left[\hat x, \left[ \hat x, \hat{\rho} \right] \right].
\end{equation} 
Such an equation appears frequently from an interaction with the environment in cases where friction is negligible. The strength of the coupling is given by the parameter $D$. The first term in (50) would lead to unitary spreading, whereas the second term would lead to nonunitary localisation. The time scale for decoherence is:
\begin{equation}
t_0 =\sqrt{\frac{m \hbar}{D}}.
\end{equation}
If we change to the Weyl-Wigner representation of quantum mechanics, we conclude that the corresponding Wigner function satisfies the Fokker-Planck equation:
\begin{equation}
\frac{dF}{dt} = - \frac{p}{m} \frac{\partial F}{\partial x} + \frac{D}{2} \frac{\partial^2 F}{\partial p^2}.
\end{equation}
If decoherence occurs, the general expectation is that the negative parts of the Wigner function will gradually be smoothed out in the course of time. Many examples support this expectation \cite{Diosi}. 

In ref.\cite{Diosi} Diosi and Kiefer proved a much stronger statement, namely that the Wigner function becomes strictly positive after a certain finite decoherence time $t_D$. This is very different from the behaviour of the density matrix whose non-diagonal elements become zero only asymptotically. This is important because the positivity of the Wigner function is usually regarded as a necessary condition for classicality.

Another relevant aspect of this mechanism is the fact that the time $t_D$, after which the Wigner function becomes strictly positive, is independent of the initial Wigner distribution. 

The results of that paper were later generalized in ref.\cite{Almeida}, where a Markovian master equation,
\begin{equation}
\frac{d \hat{\rho}}{dt} = - \frac{i}{\hbar} \left[\hat H, \hat{\rho} \right] - \frac{1}{2 \hbar} \sum_j \left( 2 \hat L_j \hat{\rho} \hat L_j^{\dagger} - L_j^{\dagger} \hat L_j  \hat{\rho} -  \hat{\rho} L_j^{\dagger} \hat L_j  \right),
\end{equation}
was considered with a set of Lindblad operators $\left\{\hat L_j \right\}$ linear in the Brownian particle's position and momentum and representing the interaction with the environment. The conclusions were identical: the Wigner function becomes strictly positive after a finite time irrespective of the initial condition.

Let us rederive the results of \cite{Diosi} using the formalism of section 2. The Green's function for the system is the smooth function (for any $t>0$) $K_t \left(z|z' \right)$, solution of the Fokker-Planck equation (52) with initial condition $K_0 \left(z|z' \right) = \delta (z-z')$. We can then write:
\begin{equation}
F(z,t) = \int dz'  \hspace{0.2 cm} K_t \left(z|z' \right) F_0 (z'),
\end{equation}
where $F_0 (z) = F(z,0)$ is the initial Wigner distribution. A straightforward calculation leads to the following solution. The propagator $K_t (z|z')$ is of the form (33), with $n=1$, $z_0^T = \left(x'+ \frac{p't}{m}, p' \right)$ and:
\begin{equation}
A(t) = \left(
\begin{array}{c c}
- \frac{6 m^2}{D t^3} & - \frac{3 m}{D t^2}\\
& \\
- \frac{3 m}{D t^2}  & - \frac{2}{D t}
\end{array}
\right).
\end{equation}
Since the propagator is of the form $K_t (z|z') = \Omega_t (z-z_0)$, we may, alternatively, rewrite (54) in the form:
\begin{equation}
F(z,t)=  \left( \Omega_t \star F_0 \right) \left(z_{-t} \right),
\end{equation}
where $z_{-t} = \left(x +  \frac{pt}{m}, p \right)$. From (33), we know that $\eta$ beleongs to the Wigner spectrum of $\Omega_t (z)$ iff the matrix, 
\begin{equation}
B_t = A_t^{-1} + i \eta J = \left(
\begin{array}{c c}
- \frac{2Dt^3}{3 m^2} & \frac{Dt^2}{m} - i \eta\\
& \\
\frac{Dt^2}{m} + i \eta & -2 Dt 
\end{array}
\right),
\end{equation}
is non-negative. We conclude that the Wigner spectrum of $\Omega_t (z)$ is:
\begin{equation}
{\cal W}  \left( \Omega_t \right) = \left[ - \frac{Dt^2}{m \sqrt 3} , \frac{Dt^2}{m \sqrt 3} \right].
\end{equation}
It follows that $\Omega_t (z)$ becomes a Wigner function once $t \ge t_D$, where:
\begin{equation}
t_D = \sqrt[4]{\frac{3 m^2 \hbar^2}{D^2}}.
\end{equation}
The Wigner function $F(z,t)$ (for $t \ge t_D$) is then the convolution of a $0$-positive Wigner function $\Omega_t (z)$ and another Wigner function $F_0(z)$ (cf(56)). From the analysis in section 2, this means that $F(z,t)$ is non-negative from $t_D$ onwards. Eq.(59) is the result obtained by Diosi and Kiefer. Notice that $t_D$ is independent of the initial Wigner distribution $F_0 (z)$.

\vspace{0.5 cm}
\noindent
How general is this mechanism? The reason why $t_D$ is independent of $F_0 (z)$ stems from the fact that it is the propagator that enters the space ${\cal F}$ of Wigner distributions at $t= t_D$ and this only depends on the dynamics of the system, i.e. on the master equation. Thus, suppose that the Wigner function at time $t \ge 0$ is given by eq.(56), where $z_t$ is some trajectory such that $z_0 =z$ and $ \Omega_t (z)$ is a propagator. Assuming that the set of conditions derived by Narcowich and analyzed in section 2 are both necessary and sufficient, we may define the set of propagators ${\cal K}$ as follows. We say that $\Omega_t (z)$ belongs to ${\cal K}$ if (i) $\Omega_t (z)$ is smooth for $t>0$, (ii) $\Omega_0 (z) = \delta (z)$, (iii) $\Omega_t$ satisfies the semi-group rule $ \left( \Omega_t \star \Omega_{t'} \right) (z) = \Omega_{t+t' } (z)$, for $t,t'>0$ and, finally, (iv) $0 \in {\cal W} \left( \Omega_t \right)$, or $\pm 2\hbar \in {\cal W} \left( \Omega_t \right)$. The latter conditions ensure that the convolution of $\Omega_t$ with an initial Wigner distribution $F_0(z)$ yield another Wigner distribution. The propagator is the solution of some master equation with initial condition (ii). Suppose that at $t_D$ the propagator becomes a Wigner function, i.e. $\hbar \in {\cal W} \left( \Omega_t \right)$ for $t \ge t_D$. Then, concomitantly the Wigner function $F(z,t)$ becomes strictly positive. From this construction it is clear that $t_D$ cannot depend upon the initial distribution $F_0 (z)$.

\vspace{0.5 cm}
\noindent
What we propose is the following. Instead of trying to derive kinetic equations for open Markowian systems, which is notoriously difficult for non-quadratic interactions (the only ones considered in refs.\cite{Diosi}, \cite{Almeida}), we may instead look for propagators in the set ${\cal K}$ which, at a given time, enter the set ${\cal F}$. We may even consider situations where the propagator enters ${\cal F}$ at time $t_C$ and leaves it at a later time $t_Q$. This is presumably what happens in ref.\cite{Antunes}. This is the programme of a future work \cite{Dias4}.

\vspace{1 cm}

\begin{center}

{\large{{\bf Acknowledgments}}} 

\end{center}

\vspace{0.3 cm}
\noindent
This work was partially supported by the grants POCTI/MAT/45306/2002 and POCTI/FNU/49543/2002.


\begin{thebibliography}{10}

\bibitem{Wigner} E.Wigner, Phys. Rev. 40 (1932) 749.

\bibitem{Lee} H.W.Lee, Phys. Rep. 259 (1995) 147.

\bibitem{Weyl} H.Weyl, Z. Phys. 46 (1927) 1.

\bibitem{Groenewold} H.Groenewold, Physica 12 (1946) 405.

\bibitem{Moyal} J.Moyal, Proc. Camb. Phil. Soc. 45 (1949) 99.

\bibitem{Baker} G.Baker, Phys. Rev. 109 (1958) 2198.

\bibitem{Carruthers} P.Carruthers, F.Zachariasen, Rev. Mod. Phys. 55 (1983) 24. 

\bibitem{Balazs} N.Balazs, B.Jennings, Phys. Rep. 104 (1984) 347.
) 677.

\bibitem{Hillery} M.Hillery {\it et al.}, Phys. Rep. 106 (1984) 121.

\bibitem{Fairlie} D.B.Fairlie, e-print: hep-th/9806198.

\bibitem{Tatarskii} V.I.Tatarskii, Sov. Phys. Usp. 26 (1983) 311. 

\bibitem{Flato} F. Bayen {\it et al.}, Ann. Phys. 111 (1978) 61; Annals of Physics 110 (1978) 111.

\bibitem{Dias1} N.C.Dias, J.N.Prata, Ann. Phys. 313 (2004) 110.

\bibitem{Dias2} N.C.Dias, J.N.Prata, Ann. Phys. 311 (2004) 120.

\bibitem{Dias3} N.C.Dias, J.N.Prata, J. Math. Phys. 43 (2002) 4602.

\bibitem{Dias4} N.C.Dias, J.N.Prata, in preparation.

\bibitem{Antunes} N.D.Antunes {\it et al.}, quant-ph/0101039.

\bibitem{Diosi} L.Diosi, C.Kiefer, J. Phys. A 35 (2002) 2675.

\bibitem{Almeida} O.Brodier, A.M.Ozorio de Almeida, Phys. Rev. E 69 (2004) 016204.

\bibitem{Walton} S.Kryukov, M.A.Walton, quant-ph/0412007, to appear in Ann. Phys.

\bibitem{Bracken} J.C.Wood, A.J.Bracken, quant-ph/0411028.

\bibitem{Werner} T.Br\"ocker, R.F.Werner, J. Math. Phys. 36 (1995) 62.

\bibitem{Kastler} D.Kastler, Commun. Math. Phys. 1 (1965) 14.

\bibitem{Loupias} G.Loupias, S.Miracle-Sole, Ann. Inst. H. Poincar\'e A 6 (1967) 39.

\bibitem{Narcowich1} F.J.Narcowich, J. Math. Phys. 29 (1988) 2036.

\bibitem{Narcowich2} F.J.Narcowich, J. Math. Phys. 30 (1989) 2565.

\end{thebibliography}
\end{document}